# Survey and Analysis of IoT Operating Systems: A Comparative Study on the Effectiveness and Acquisition Time of Open Source Digital Forensics Tools

Jeffrey Fairbanks, GCFA, GDAT
*Department of Computer Science*
*Boise State University*
Boise, Idaho, USA
JeffreyFairbanks@u.boisestate.edu

Md Mashrur Arifin
*Department of Computer Science*
*Boise State University*
Boise, Idaho, USA
MdmashrurArifin@u.boisestate.edu

Sadia Afreen
*Department of Computer Science*
*Boise State University*
Boise, Idaho, USA
SadiaAfreen@u.boisestate.edu

Alex Curtis
*Department of Computer Science*
*Boise State University*
Boise, Idaho, USA
AlexCurtis@u.boisestate.edu

*Abstract*—The main goal of this research project is to evaluate the effectiveness and speed of open-source forensic tools for digital evidence collecting from various Internet-of-Things (IoT) devices. The project will create and configure many IoT environments, across popular IoT operating systems, and run common forensics tasks in order to accomplish this goal. To validate these forensic analysis operations, a variety of open-source forensic tools covering four standard digital forensics tasks. These tasks will be utilized across each sample IoT operating system and will have its time spent on record carefully tracked down and examined, allowing for a thorough evaluation of the effectiveness and speed for performing forensics on each type of IoT device. The research also aims to offer recommendations to IoT security experts and digital forensic practitioners about the most efficient open-source tools for forensic investigations with IoT devices while maintaining the integrity of gathered evidence and identifying challenges that exist with these new device types. The results will be shared widely and well-documented in order to provide significant contributions to the field of internet-of-things device makers and digital forensics.

## I. Introduction and Motivation

Internet of Things (IoT) devices have become prolific in modern life. These devices help and entertain us and improve everyday life. These devices can contain a wealth of information for a Digital Forensic investigator that knows how to get information off of them [23]. The motivation is to understand how a Digital Investigator would approach an investigation that involved one of these devices. Cybersecurity statistics indicate that there are 2,200 cyber attacks per day, with a cyber attack happening every 39 seconds on average. In the US, a data breach costs an average of $9.44M, and cybercrime is predicted to cost $8 trillion by the end of 2023 [1].

The critical need to determine which data acquisition tools are the most effective, economical, and versatile in the complex world of the Internet of Things (IoT) is what motivates this research. Within the realms of digital forensics and cybersecurity, prompt and accurate data collection is crucial, particularly in situations where there are numerous firmware versions and distributed IoT architectures. These features are intended to minimize downtime, enable secure infrastructure restoration, and enable quick responses to security threats and outages. By executing timed data acquisitions on several systems, this study seeks to provide a standard for data capture and thoroughly evaluate different tools. The main goal is to identify the tool that performs the best in terms of time efficiency, computational cost-effectiveness, and adaptability for a variety of IoT devices. The primary focus of the research is on IoT devices because they have become part of many different applications inn modern day-to-day life, making their security crucial [26]. The results of this study will give cybersecurity experts useful information to help them choose the most appropriate tools for effective data collection and investigation, boosting the robustness and efficiency of forensic measures in Internet of Things environments, protecting vital infrastructure, and analyzing the incident in most efficient manner.

The economies of scale that come with maintaining and ensuring stability in Internet of Things (IoT) systems are also a major factor in ensuring their well-being. By 2025, the IoT sector is expected to have an 11 trillion dollar economic impact, according to projections made in 2019. 75 billion IoT devices are expected to reach prospective customers worldwide by 2025, according to a 2016 Statista Research Department analysis [11]. By 2023, the total amount spent globally on IoT would have reached 1.1 trillion dollars.

In order to fulfill the correct and expected outcomes within this research, research guidelines were constructed. They are bulleted below:

1. Evaluate the effectiveness and speed of open-source forensic tools for digital evidence collecting from various Internet-of-Things (IoT) devices.
2. Create and configure popular IoT operating systems and run common forensics tasks in order to accomplish this goal.
3. The research also aims to offer recommendations to IoT security experts and digital forensic practitioners about the most efficient open-source tools for forensic investigations

Fundamental research questions were also crafted with the intention of giving clear outcomes. First, when used on multiple IoT devices, which tool has the most time efficiency? The main focus of this study is on each tool's capacity to obtain pertinent data from the IoT system promptly. The question of which tool provides the most computationally economical solution is also scoped to be answered through this research. Digital forensics analysts entrusted with data extraction and IoT device forensics may encounter difficulties with solutions that demand a large amount of processing power from the CPU or RAM. In light of the second study question, it's critical to ascertain whether there is a single optimal tool that performs better than all others across a range of IoT devices. Finally, certain tools might be better suited for a particular device. Even though some of these tools are exclusive to a certain firmware or IoT device, the researchers believe it is still vital to showcase these resources. With the use of this data, analysts will be better able to determine whether to maintain an environment that supports several IoT platforms or to build tools that are only required for a particular IoT design.

## II. Prior Work

In surveying the existing body of literature, this section delves into the relevant works that have contributed to the understanding of the intersection between IoT and digital forensics, providing a comprehensive overview of key studies and advancements in the field. In the realm of the Internet of Things (IoT), there is widespread discussion concerning the current state of IoT. Several surveys have explored the general landscape of digital forensics in IoT, as discussed below. However, it is important to note that none of these surveys have conducted practical assessments of the commonly used open-source digital forensics tools in traditional digital forensics tasks. The following section outlines the main works in this area.

In the paper, *A Survey on the Internet of Things (IoT) Forensics: Challenges, Approaches, and Open Issues*, [25] a survey of digital forensics theory is given to encapsulate the wide range of IoT devices. This paper correctly points out that, while IoT data presents a valuable evidentiary resource, as a results of the wide adoption of these edge devices, forensic experts encounter a range of challenges. These challenges highlighted in this paper stem from extensive diversity of IoT devices and non-standard data formats. The study offers a comprehensive examination of historical and contemporary theoretical models within the field of digital forensics. While the authors do a great at highlighting and recommending a digital forensics theory, it does not test the tools across a wide range of IoT devices to give recommendations on tools effectiveness, and instead focuses on theoretical digital forensics frameworks.

In the paper, *Internet of Things(IoT) digital forensic investigation model: Top-down forensic approach methodology*, the authors seeks to determine an optimal approach by creating an innovative model for handling investigation scenarios in digital forensics. [24] The authors note that while previous research has presented models for identifying forensic objects in investigations, there has been a lack of thorough testing to validate the efficacy of these approaches. In the present study, an integrated model is proposed. This paper centers on examining frameworks within the domain of digital forensics and their applicability to IoT devices. Notably, it refrains from empirically evaluating the efficacy of existing open-source tools commonly employed in traditional digital forensic tasks.

In the paper, *A Survey on Digital Forensics in Internet of Things*, the researchers condense projects between 2010 and 2018 and give an updated systematic approach to IoT. [16] The researchers expand on the previous work by creating what they describe as a 3d model of the current landscape of IoT, in hopes to help standardize the space. The contribution of this research helps in proposing insightful recommendations to spur future research. By considering the researchers' suggestions for future studies and examining the current state of IoT digital research, papers like this one can better grasp ongoing challenges and steer clear of potential pitfalls.

In the paper, *IoT Forensics: An Overview of the Current Issues and Challenges*, the authors explored challenges faced by digital forensic investigators, spanning issues like uncertain data location, diverse devices, data volatility, and inadequate forensic tools. [18] The main area of interest for this paper is on the tools. However, the paper posited by Janarthanan et al. primarily underscores potential challenges in investigations without substantiating the validity of these assertions. The researchers underscore potential issues associated with traditional digital forensics tools, such as variability among IoT devices, the presence of proprietary hardware and software, and the susceptibility of data to updates, modifications, or loss. However, these assertions lack empirical validation through rigorous testing of the tools and comparative analysis of outputs.

In the paper, *Forensic State Acquisition from Internet of Things (FSAIoT): A General Framework and Practical Approach for IoT Forensics*, the authors talk about the FSAIoT framework proof of concept implementation that is proposed through their research. [21] In their publication, the authors observe that by harnessing the acquisition capabilities of IoT devices, a lucid representation of transpired events can be achieved. The proof of concept embedded within the open-source tools shifts the spotlight onto IoT digital forensics as the primary focus of the tool. This paper outlines itself within the corpus of related works by innovatively introducing a tool explicitly designed for IoT digital forensics—a rarity, considering that numerous tools with relevance to IoT often transition into commercial domains due to the very broad scope in this evolving field.

Upon investigation of related work in this domain, it became evident that there is a necessity for research to assess the validity of presently employed open-source tools in traditional digital forensics and to juxtapose the obtained results against those produced by various other open-source tools.

## III. Methodology

This study aims to evaluate digital forensics tools employed on Internet of Things (IoT) Operating Systems, with a focus on identifying tools that exhibit high information richness, computational efficiency, and time effectiveness.

The research establishes a benchmark framework to systematically collect data from diverse IoT Operating Systems, contributing to the fulfillment of the objective. Each selected tool undergoes timed acquisitions on various IoT devices, generating benchmark data that is subsequently used to construct a matrix detailing acquisition times across different platforms. These averages serve as the basis for determining the overall performance superiority of each tool. The assessment is further nuanced by considering each tool's limitations in data collection, recognizing the necessity for Digital Forensics Analysts to possess versatile tools capable of effectively handling data from an array of IoT devices. This research endeavors to provide a valuable resource mitigating challenges arising from the diverse firmware and operating systems prevalent in various IoT devices. Additionally, the study accounts for the specific restrictions each tool encounters when dealing with distinct types of IoT devices.

1. **Standardize Data Collection:** Define a standardized protocol for data collection across different IoT systems.
2. **Tool Deployment and Timed Acquisitions:** Apply each selected tool to conduct timed acquisitions of benchmark data on each IoT device.
3. **Matrix Creation:** Develop a comprehensive matrix detailing acquisition times for each tool across diverse IoT platforms.
4. **Performance Evaluation:** Assess overall tool performance by calculating average acquisition times.
5. **Data Collection Restrictions:** Document and consider the restrictions imposed by each tool on the types of data it can effectively gather from IoT devices.

Through these outlined approaches, the study endeavors to provide a systematic and comprehensive evaluation of digital forensics tools in the context of IoT Operating Systems, offering insights into their comparative efficiency and suitability for diverse IoT environments. Figure 1 demonstrates this process flow for the research methodology.

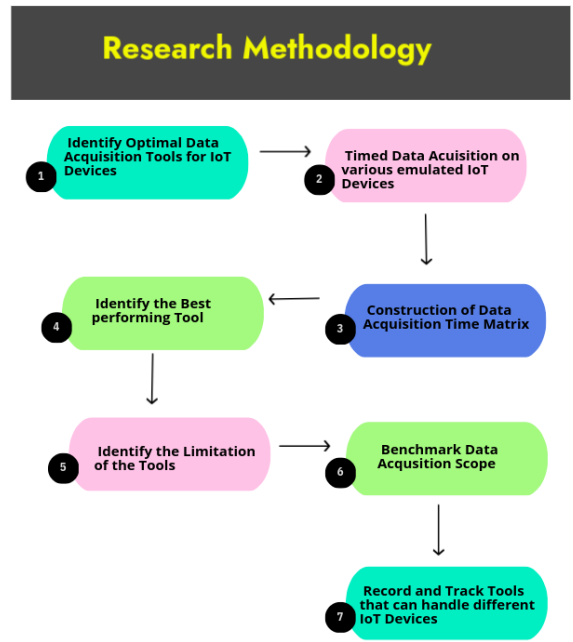

Figure 1. Research Methodology

Within the scope of this project, a meticulous selection of Internet of Things (IoT) operating systems is undertaken to encapsulate the diverse spectrum of IoT devices. Subsequently, tools aligned with traditional digital forensics methodologies, as commonly applied in investigations pertaining to conventional end-user operating systems, such as Windows OS, are chosen.

Once the IoT operating systems and tools are identified, a systematic evaluation is completed on each tool as they undergo testing on every selected IoT operating system. Comprehensive data, encompassing metrics such as time elapsed, CPU utilization, and various other indicators, is also meticulously collected through this process. Following this evaluative process, a comparative analysis is conducted to discern and interpret the numerical outcomes. The results are then synthesized to provide a comprehensible representation of the multifaceted indicators derived from the evaluation. Ultimately, the research offers informed recommendations and discourse on the efficacy of each tool when applied in the context of IoT operating systems, a scenario not conventionally considered within their typical use cases.

### A. IoT Operating Systems

The strategic decision to leverage IoT Operating Systems rather than individual systems is an important point to highlight. Through a discerning selection of diverse IoT Operating Systems, this research attains a comprehensive perspective on the IoT landscape while streamlining the need to directly profile numerous devices. This approach facilitates robust testing of open-source traditional digital forensics tools, yielding distinct and nuanced outcomes across the expansive IoT domain.

In this study, the selection of well-known IoT operating systems has been made, including WindowsIoT, Balena, Fortinet, and Contiki. These choices were based on the candidates' well-established reputations in the IoT community and their shown importance to the field of IoT forensics. It is interesting to note that these operating systems are well-known solutions that have been widely used in the IoT field. This deliberate selection of tools makes sure that the study takes advantage of the knowledge and features provided by tools that are often used by IoT forensics professionals, strengthening the validity and practical relevance of the conclusions.

*1) Balena Raspberry IOT OS:* Popular for IoT development and deployment, Balena is renowned for its scalability and flexibility. It provides a container-based methodology that makes it simple to scale and manage apps across a variety of IoT devices. It was essential to the study because of its capacity to reliably support the research, monitor the health of large fleets of devices, and distribute software upgrades to them with ease. [7]

*2) Fortinet IOT OS:* Well-known for its strong security features is Fortinet. It provides IoT environments with a high degree of network security and threat prevention as a well-known cybersecurity solution. Fortinet's forensics tools were especially useful in the investigation for examining security incidents and vulnerabilities in Internet of Things platforms. Its features for preventing and detecting intrusions are especially important. [12]

*3) Windows IoT OS:* The familiarity and ease of use of Windows IoT are its defining characteristics. It makes Internet of Things development accessible to a broad spectrum of developers by offering a Windows-based environment. For individuals who are familiar with Microsoft's environment, this platform makes it easy to integrate IoT forensics procedures with Windows-based tools and applications. [19]

*4) Contiki IOT OS:* Contiki is renowned for being both energy-efficient and lightweight. Its ability to function on IoT devices with limited resources is essential for those with low processor and memory capacities. Notable features of Contiki include its compatibility with battery-operated IoT devices and support for low-power wireless communication protocols, which make it indispensable for particular IoT situations in the study. [3]

*B. Forensic Tools*

In order to perform the forensic tasks against the operating systems listed above, common forensics tools were used to understand and study how they work with these types of devices and systems. All of the tools selected for this study are open-source and readily accessible.

All of these tools are well-known solutions that have been widely used in the traditional forensics field, such as through investigations on Windows OS and Linux OS, rather than being created specifically on IoT devices and operating systems.

This deliberate selection of tools, allows this research to test the effectiveness and validity of these tools on IoT infrastructure. This research aims to assess these tools for identifying the most time-efficient, information-rich, and computationally cost-effective option within the context of IoT systems. The resulting data is then used to construct a matrix that records the acquisition times for each tool across different systems. The tool that demonstrates the best overall performance is identified based on these averages. Additionally, limitations pertaining to the types of data certain tools can collect are factored into the final score of each tool. Having a "jack-of-all-trades" type tool in a Digital Forensics Analyst's tool chest is extremely valuable in ensuring that all data can be collected, sifted, and understood in a timely and effective manner across the wide range of different IoT Operating Systems. The tool that can mitigate the challenges of differing operating systems is widely valuable and is recorded and tracked through this research.

*1) FTK Imager:* FTK Imager's[1] strong features and track record of dependability are the reasons this tool was selected for forensic image acquisition in IoT OS forensics. With its extensive support for a multitude of file systems, FTK Imager is an extremely efficient tool for obtaining forensic images from a variety of operating systems that are frequently found in Internet of Things devices. This tool is able to gather and evaluate digital evidence from Internet of Things operating systems with greater efficiency, as well as due to the tool's intuitive interface and strong imaging capabilities led to its usage. FTK Imager is a flexible solution that tackles the particular difficulties presented by IoT environments because of its capacity to generate both logical and physical images in addition to its ability to record volatile data. FTK Imager is considered to be a reliable option in the field of IoT OS forensics because of its track record of preserving data integrity and its compatibility with a wide range of storage media. [6]

*2) DD:* For forensic image acquisition in IoT OS forensics, the `dd` (data duplicator) command [2] was selected because of its ease of use and adaptability. The majority of Unix-like operating systems offer the `dd` command pre-installed on the operating system, which makes it ubiquitous. Its simple syntax and user-friendliness make it a practical option for getting forensic images from different IoT device operating systems. `dd` gives us the ability to make exact replicas of physical and logical storage media, offering a thorough method for acquiring digital evidence. Because of its flexible and customizable command-line interface, it can be used with a variety of scenarios and storage media types that are frequently found in Internet of Things environments. `dd` is a sensible option for the IoT OS forensic investigations because of its dependability and the guarantee it offers in preserving data integrity throughout the forensic imaging procedure. [4]

---

[1] See: https://www.exterro.com/ftk-imager
[2] See: https://pubs.opengroup.org/onlinepubs/9699919799/utilities/dd.html

*3) Volatility:* An open source tool built in Python, Volatility[3] offers advanced analysis of volatile data from a computer's memory dump. It is broadly compatible with various operating systems like Windows, Linux, and Mac OS. This versatility is crucial and tested in this study how it will operate against the lesser-known IoT operating systems of this study. The tool operates by parsing and interpreting memory dumps, enabling forensic analysts to uncover hidden or obfuscated data that might not be evident through traditional disk forensics. Volatility distinguishes itself with a plugin-based architecture, fostering an environment for continuous expansion and specialization, from community-provided plugins which are distributed through Github. This framework empowers users to develop custom plugins tailored to specific investigative needs or novel forensic scenarios, thus keeping pace with the rapidly evolving digital landscape. [8]

*4) SleuthKit:* The Sleuth Kit[4] is a powerful tool used in the field of digital forensics. This toolkit contains a variety of Unix and Windows command line utilities used for extracting data from disk drives and other storage mediums. At its core, Sleuth Kit excels in the analysis of file system structures. It can be used to recover deleted files, examine file system metadata, and identify evidence of file tampering or deletion. This capability is crucial in digital forensic investigations, where understanding the history and state of files is essential in uncovering activities on a digital device. Many of the commands were used in this set for both metadata layer and filesystem layer analysis. These commands work together cohesively for example using `fls` to search the filesystem and find appropriate inodes, and then using `istat` to investigate metadata related to the files. SleuthKit supports all common filesystems both legacy and modern. These can be selected at run time, which could expect to offer strong compatibility across all the IoT systems tested. [14]

## C. Forensic Tasks

The foundational premise of this investigation lies in the examination of commonplace scenarios encountered by digital forensics investigators, where the imperative is to extract evidence from Internet of Things (IoT) devices. Within each scenario, the requisite activities of a forensic investigator are condensed into a concise set of four fundamental tasks. The analysis of this research will focus on delineating the methodologies through which an investigator can proficiently execute these tasks across four prevalent IoT operating systems.

*1) Image Acquisition:* Capturing comprehensive images of the storage media within IoT devices, image acquisition tools such as FTK (Forensic Toolkit) and DD (Disk Dump) are employed. Renowned for its user-friendly interface, FTK excels in swiftly acquiring disk images, preserving file properties, and accommodating various file systems. By comparison, `dd` is a command-line utility proficient in generating bitwise copies of storage media. The synergistic utilization of these tools facilitates meticulous data collection from IoT devices, ensuring both data integrity and preservation.

When using FTK Imager, the media was mounted onto the host. FTK was then used to create a disk image. *Logical Drive* was selected as the source in FTK, pointing to the mounted drive of the IoT device. Other options that were used are the "Verify images after they are created" and "Precalculate Progress Statistics". The former setting runs a second pass on the completed image once complete to compare the hash of the original source and the completed image to ensure integrity of the image. The latter setting was simply used to accurately inform researchers of progress since this imaging process can take time. Importantly, `0` was selected for the fragment size, which is recommended for Raw formats.

By selecting the verification step, FTK Imager also provides a detailed text output about the process, including the checksum values of the verification, geometry of the drive, and timing data. This output file is what was used as timing data for this portion of study.

*2) Memory Analysis:* This is an important part of IoT forensics, volatility was the best selection to fulfill this work as a result of its ability to analyze and decipher memory dumps. Numerous memory analysis methods, such as identifying processes, network connections, and registry data, are available with volatility tools. They are essential to forensic investigations because they make it possible to reconstruct the state of the device at the moment of data gathering.

*3) File System Analysis:* File system analysis is a critical component in the field of digital forensics, playing a pivotal role in the investigation of devices. It involves the examination and interpretation of file systems to recover, analyze, and present data from storage mediums. This analysis is essential for uncovering evidence that could be hidden, deleted, or manipulated. A successful analysis will provide a strong understanding of storage hierarchies and taxonomies on the tested devices including valuable information like inode offsets which can help with further investigation.

The first command performed in this section of the study is `blkls`, which is a useful tool for sector-level data recovery and is used to retrieve data from individual disk blocks. [2]

*4) Metadata Analysis:* Metadata level analysis in digital forensics refers to the examination and interpretation of metadata associated with digital files and electronic communications. Metadata is essentially data about data which can provide valuable information about the creation, modification, and usage of digital artifacts. This type of analysis is a crucial component of digital forensics investigations because it can help investigators reconstruct events, track user activities, and establish timelines of digital evidence.

Inode analysis helps with the mapping of file structures and metadata. This process helps with the comprehension of file properties and the reconstruction of file hierarchies. When used in tandem with file analysis tools like `fls`, these technologies offer extensive insights into IoT device file systems, facilitating thorough forensic analysis.

---

[3]See: https://github.com/volatilityfoundation/volatility
[4]See: https://sleuthkit.org

There are numerous tools available for metadata level analysis. The focus throughout this paper is on open source tools that are accessible to all investigators. The Sleuth Kit [14] includes three common tools, `fls` [5], `ils` [9], and `istat` [10] which are the tools focused on throughout this study.

By starting with the `fls` command, investigators can gather information about the files present on a system, including their names, sizes, timestamps, and inode numbers. This is the first command attempted in this study to identify a file to target.

As a second step, the `ils` command was used to retrieve detailed information about specific files or directories on a storage device or disk image. This can even be used for identification of deleted files.

As a third and final command within this section of the study, the `istat` command was used to pull information about an inode that was identified with the previous commands. By using `istat`, investigators can gather crucial data to reconstruct a file's history, access times, and other metadata, aiding in digital forensics examinations and investigations. The returned data helps analysts piece together the story of a file's existence and usage within the file system.

All of this information collectively can be crucial for identifying and prioritizing potential evidence during a forensic examination.

### D. Programming Language and Platforms

In the research, Python served as the primary programming language for the implementation and execution of forensics tasks. The selection of Python was deliberate, as it offered a combination of versatility and access to a rich ecosystem of libraries. It also mixes well with the Linux shell.

SANS SIFT workstation [13] is a specialized Linux distribution designed for digital forensics and incident response tasks. It provides a comprehensive set of open-source forensic tools and libraries, making it a valuable resource for forensic analysts and investigators. This workstation is available for download from the SANS institute website[5] and provides the exact environment used to run all commands and analysis of this study. This platform is freely distributed and open source.

### E. Tool Implementation and Testing

In the study, each tool was carefully experimented with for specific forensics tasks, such as memory scanning, Volatility analysis, image acquisition with tools like FTK and DD, and file system analysis with blkls and ils/inodes. The forensics jobs were also carried out on four well-known IoT operating systems: Balena, Fortinet, Windows IoT, and Contiki. To begin, the collection of IOT devices were downloaded from their respective official websites while ensuring the integrity of the acquisition process. Subsequently, these images were deployed across multiple environments, including VMware Workstation, Oracle VirtualBox, and Docker. Finally, the images of the IOT operating systems were deployed in these environments through the utilization of `dd` and FTK Imager.

---

[5]See: https://www.sans.org/tools/sift-workstation/

| IoT Operating Systems | Image Size | Image Types |
|---|---|---|
| FortiOS | 42.8 Mb | .ovf file |
| Contiki | 6.6 Gb | .vmdk, .vmx, .vmxf, .nvram |
| Balena OS | 908 Mb | .img |
| Windows IoT | 1.4Gb | Raw (.001) |

Figure 2. Size of the Image for each IOT Operating System

During the image creation process, there were a few specific challenges. In the case of virtual machines, it was not always feasible to capture all files within an image. Instead, snapshots were taken of each partition, which were denoted as `.vmdk` files in the virtual environment. Additionally, volatility tools were deployed to analyze memory dumps and Sleuth Kit was utilized for file system analysis.

The use of a virtual machine for data collecting was a distinct choice made by the researchers to keep data collection patterns consistent. Other data collections exists, such as collecting data over WAN or using device inputs [20], but these techniques are device-specific and require customized collection methods per device.

This approach enabled the project to conduct a comprehensive comparison of time and performance results. One of the objectives is to ensure the repeatability of this study; therefore, all steps are explicitly laid out and benchmarks are established to facilitate future assessments.

## IV. RESULTS AND ANALYSIS

Each section of the study's analysis is provided in detail below. Figures and screenshots are provided when notable value is available.

### A. Examination of Internet Of Things Operating System metadata

When examining the size and other metrics (Figure 2) of different IoT operating systems (OSs), it is clear that there is a significant variation in their image sizes and formats. FortiOS is a compact option, with an image size of 42.8 Mb in the `.ovf` file format.

Contrarily, Contiki exhibits a notably larger size of 6.6 Gb, comprising diverse file formats including `.vmdk`, `.vmx`, `.vmxf`, and `.nvram`.

Balena OS is positioned between these two extremes, as it has an image size of 908 Mb in the `.img` format. Windows IoT, with a 1.4 Gigabyte image size in Raw (`.001`) format, is classified as a mid-range option.

Although the IoT operating systems vary in size, it is important to note that they do have some similarities in terms of their size ranges. However, the decision between them may involve factors other than just size. Additionally, it is important to note that Fortinet's FortiOS `.ovf` file permits the adjustment of disk size during deployment, providing a significant degree of adaptability that is essential for meeting specific IoT project needs.

```
Created By AccessData® FTK® Imager 4.7.1.2

Case Information:
Acquired using: ADI4.7.1.2
Case Number: 02
Evidence Number: 02
Unique description: fgos02
Examiner: md
Notes:

--------------------------------------------------------

Information for C:\Masrur\Output.img\fgos02:

Physical Evidentiary Item (Source) Information:
[Device Info]
 Source Type: Physical
[Drive Geometry]
 Cylinders: 261
 Heads: 255
 Sectors per Track: 63
 Bytes per Sector: 512
 Sector Count: 4,194,304
[Physical Drive Information]
 Drive Interface Type: lsilogic
[Image]
 Image Type: VMWare Virtual Disk
 Source data size: 2048 MB
 Sector count:    4194304
[Computed Hashes]
 MD5 checksum:    0d79b9ee51e9f5afbc5a6652de32a767
 SHA1 checksum:   f5b6cda4bd07e3667798cc30ae1a82ee02439680

Image Information:
 Acquisition started:   Sat Nov 11 13:21:59 2023
 Acquisition finished:  Sat Nov 11 13:22:03 2023
 Segment list:
  C:\Masrur\Output.img\fgos02.001

Image Verification Results:
 Verification started:  Sat Nov 11 13:22:03 2023
 Verification finished: Sat Nov 11 13:22:09 2023
 MD5 checksum:    0d79b9ee51e9f5afbc5a6652de32a767 : verified
 SHA1 checksum:   f5b6cda4bd07e3667798cc30ae1a82ee02439680 : verified
```

Figure 3. Image Acquisition of FortiOS by FTK Imager

## B. Image Acquisition

The process of image acquisition, employing *dd* and *FTK Imager*, demonstrated consistent success throughout the entirety of the research endeavor. Both methodologies effectively fulfilled the task of imaging the Internet of Things (IoT) operating systems. The ensuing section is dedicated to a detailed examination and comparative analysis of the outcomes obtained from each tool in the context of the IoT operating systems.

In the initial stages of forensic investigation, images for Fortinet and Contiki were acquired using the `dd` tool. The images for Balena and Windows IoT were obtained through the utilization of FTK Imager. As the inquiry progressed, the methodology was expanded to include the acquisition of images for Fortinet utilizing FTK Imager (See Figure 3). The execution of this strategic change enabled the adoption of a unified methodology across multiple platforms, ensuring consistency and thoroughness in the forensic analysis. By tailoring image acquisition methods to accommodate the specific requirements and nuances of individual systems, the investigation could be conducted with greater precision and thoroughness, thereby enhancing the forensic process's robustness and dependability. The rationale for this methodology is predicated on deliberately constraining the resources within the experimental configuration and conducting a meticulous assessment of the operating system's dependability [15].

The crucial factor supporting this strategic choice is the substantial resource allocation that is necessary for particular Internet of Things (IoT) operating systems, including Windows IoT, Fortinet, and Contiki. Critical tools such as FTK imager are exclusively compatible with the Windows operating system. The objective is to optimize the functionality and efficiency of the experimental environment while adhering to these constraints. Ensuring the compatibility of resources and operating systems as a top priority facilitates a streamlined and efficient research and development process with respect to IoT technologies.

Figure 4. Image Acquisition of FortiOS by dd

The dd command outputs in Figure 5 demonstrate significant disparities in performance between FortiOS and Contiki OS. Although there are differences in data copy size, system time, CPU utilization, and file system inputs/outputs between the two systems, FortiOS typically shows lower CPU usage, shorter system times, and a constant maximum resident set size of 6272 kbytes. Conversely, Contiki OS typically demonstrates elevated CPU utilization, extended system duration, and a marginally larger maximum resident set size of 6400 Kb during a single operation. These differences indicate that FortiOS may provide more reliable and effective performance in various `dd` operations, making it potentially more desirable in situations where resource efficiency is crucial. Nevertheless, the most suitable option between the two would rely on the specific requirements of the use case.

The table (Table VII) presents a comparison of different image acquisition parameters using FTK Imager for Fortinet, Balena, and Windows IoT on various disks. Fortinet employs FG0s2 and FG0s3 as disk segments, whereas Balena and Windows IoT have distinct designations such as BalenaRaspberry.001, Data-f.002, efi-e.001, and mainos. All disks have a uniform sector size of 512 bytes. However, there are variations in the types of images used by different platforms. Fortinet utilizes VMWare Virtual Disk, Balena chooses Raw (dd), and Windows IoT employs Virtual Disk for all three disks. Fortinet Disk2 has the largest source data size, measuring 30720 MB, whereas Balena's is the smallest, measuring 908 MB. The sector counts differ, and Fortinet Disk2 has the highest count at 62914560. The acquisition times demonstrate variations in efficiency, with Fortinet Disk1 being the most rapid at 4 seconds and Windows IoT Disk2 taking the longest duration of 2 minutes and 16 seconds. This thorough comparison provides a clear understanding of the various acquisition parameters and performance metrics, offering valuable guidance for selecting the most appropriate configuration based on specific forensic needs.

| DD Parameters | FortiOS | | Contiki OS | | | | | |
|---|---|---|---|---|---|---|---|---|
| records in | 16+1 | 12+1 | 299+1 | 446+1 | 388+1 | 294+1 | 153+1 | 0+1 |
| records out | 16+1 | 12+1 | 299+1 | 446+1 | 388+1 | 294+1 | 153+1 | 0+1 |
| Byte Copied | 70451200 bytes (70 MB, 67 MiB) copied, 0.252733 s, 279 MB/s | 53608448 bytes (54 MB, 51 MiB) copied, 0.268034 s, 200 MB/s | 1254293504 bytes (1.3 GB, 1.2 GiB) copied, 3.70457 s, 339 MB/s | 1873870848 bytes (1.9 GB, 1.7 GiB) copied, 6.01379 s, 312 MB/s | 1627848704 bytes (1.6 GB, 1.5 GiB) copied, 4.84199 s, 336 MB/s | 1234173952 bytes (1.2 GB, 1.1 GiB) copied, 4.10635 s, 301 MB/s | 643694592 bytes (644 MB, 614 MiB) copied, 1.91156 s, 337 MB/s | 65536 bytes (66 kB, 64 KiB) copied, 0.00132197 s, 49.6 MB/s |
| System time (seconds) | 0.09 | 0.09 | 1.89 | 2.86 | 2.48 | 1.97 | 0.98 | 0 |
| Percent of CPU this job got | 35% | 35% | 51% | 47% | 51% | 48% | 51% | 66% |
| Elapsed (wall clock) time (h:mm:ss or m:ss) | 0:00.27 | 0:00.27 | 0:03.70 | 0:06.01 | 0:04.84 | 0:04.10 | 0:01.91 | 0:00.00 |
| Maximum resident set size (kbytes) | 6272 | 6272 | 6272 | 6272 | 6272 | 6272 | 6400 | 2304 |
| Minor (reclaiming a frame) page faults | 1121 | 1121 | 1121 | 1121 | 1120 | 1119 | 1122 | 113 |
| Voluntary context switches | 375 | 375 | 8581 | 12911 | 10575 | 8274 | 4260 | 2 |
| File system inputs | 104592 | 104592 | 2449768 | 3659888 | 3179360 | 2410472 | 1257184 | 96 |
| File system outputs | 104704 | 104704 | 2449792 | 3659904 | 3179392 | 2410496 | 1257216 | 128 |
| Page size (bytes) | 4096 | 4096 | 4096 | 4096 | 4096 | 4096 | 4096 | 4096 |

Figure 5. Output from dd: FortiOS vs Contiki

Upon analyzing the image acquisition parameters of dd and FTK Imager on various IoT operating systems, it is evident that FTK Imager is the superior tool for this specific task. When examining the given parameters, FTK Imager shows impressive acquisition times, particularly with Win-IoT Disk1 and Win-IoT Disk3, where it significantly outperforms dd. The efficiency of FTK Imager is especially apparent in its capability to handle various disk segments and data sizes while maintaining reasonable acquisition times. Moreover, FTK Imager offers an intuitive interface and a range of specialized functionalities designed specifically for digital forensics activities, rendering it a pragmatic option for extracting images from IoT devices. The efficiency of an image acquisition tool can be evaluated based on parameters such as acquisition time, bytes per sector, and sector count. FTK Imager demonstrates strong performance in these metrics. Hence, considering the provided data, FTK Imager is suggested for IoT image acquisition owing to its commendable performance and aptness for digital forensics tasks.

*C. Memory Scanning*

The Memory Scanning segment of the investigation yielded diverse and noteworthy findings. Primarily, Volatility exhibited limitations in effectively retrieving pertinent information for forensic examination. Across the various Internet of Things (IoT) operating systems, Volatility encountered difficulties in extracting image information from all instances. This deficiency is sub-optimal for forensic examiners seeking comprehensive insights into the memory content within IoT operating systems. As expounded in the subsequent File Layer Analysis and Metadata layer sections, alternative tools demonstrated greater efficacy in locating image information within the IoT operating system, whereas volatility encountered challenges in this specific data extraction task. Furthermore, the encountered limitations with volatility underscore the nuanced challenges in forensic memory analysis within the dynamic landscape of IoT operating systems. The distinctive characteristics and resource constraints inherent in these environments pose obstacles to conventional memory scanning techniques. This divergence in efficacy prompts a critical evaluation of the adaptability and robustness of forensic tools in addressing the unique intricacies associated with IoT memory forensics.

While Volatility proved inadequate in gathering pertinent information from any of the IoT operating systems, an evaluation of each of its plugins is undertaken to gauge their effectiveness concerning CPU utilization, execution time, and other relevant factors. The ensuing tables provide a comparative analysis of these plugins, shedding light on their respective performance metrics. The first metric comparison that is expounded on is the `imageInfo` Volatility plugin. This plugin aims to collect information about the profile that will be used in the respective Volatility plugins.

The following tables present key performance metrics for different operating systems (OS) investigated in the study: FortiOS, EFI, Data-f, Contiki, and Balena. The metrics include Mean Elapsed Time (seconds), Mean Percent of CPU used, Mean File system inputs, Mean Involuntary Context Switches, Mean Voluntary Context Switches, and Major I/O Page Faults.

As can be seen in Table II, FortiOS demonstrated the longest mean elapsed time at 420.3 seconds, while EFI and Data-f exhibited considerably shorter duration at 23.0 and 22.0 seconds, respectively. Contiki and Balena fell in between, with Contiki displaying a mean elapsed time of 153.2 seconds, and Balena recording 465.6 seconds. All operating systems, regardless of the OS type, consistently utilized approximately 99.0% of the CPU, indicating high processing demands across the board. This is not surprising as the Volatility plugin tries to vigorously find the layout and image information of each of the operating systems. FortiOS again recorded the highest mean file system inputs at 562,851,488.0, significantly surpassing the other operating systems. Contiki and Balena showed intermediate values, while EFI and Data-f exhibited the lowest mean file system inputs. The `imageInfo` volatility command ran with the highest time overall between all of the commands inspected and also had far more numerous file system inputs than the rest of the commands. As was previously stated, Volatility proved inadequate in gathering pertinent information from any of the IoT operating systems, but the metrics differing between each are still interesting.

The next command used, `pstree`, also gave some interesting results. The results are referenced in Table III. Comparing `pstree` with `imageinfo` demonstrates that both commands have very high mean file system inputs when compared to the rest of the commands. However, the time taken to complete the objective is far lower than that which was seen with the `imageinfo` command. When it comes to the rest of the statistics gathered in this command, the IOT Operating Systems all stick closely to each other.

In the command `connscan`, referenced in Table (Table IV), there begins to be a breakaway in the various metrics collected between each of the IOT Operating Systems. The IOT systems FortiOS, EFI, and Data-f conduct zero mean file system inputs and effective zero context switches with zero I/O page faults. The other IOT Systems, Mainos, Contiki, and Balena chug along with outcomes that make sense with the data provided. Each progressively runs longer based on the mean file system inputs.

The command `pslist`, referenced in Table (Table V), confirms much of what has been depicted in the previous figures with Mainos and Contiki taking up much of the Mean File system inputs, elapsed time, and Mean Voluntary Context Switches.

The command `hivescan`, referenced in Table (Table VI) is the next command that was extracted and investigated. In this command, it is clear that Contiki spent much computational time performing the hive scan. however, the time taken to complete the objective was on par with the other front runners Balena and Mainos.

This comprehensive analysis of system performance metrics provides valuable insights into the comparative efficiency and resource utilization of the examined operating systems under various conditions. The observed variations highlight the nuanced performance characteristics that may influence the selection of an operating system based on specific requirements and priorities.

### D. File System Analysis

File System Analysis constitutes a crucial aspect of digital forensics, especially when dealing with investigations involving multiple IoT operating systems. Each OS typically utilizes a distinct file system, complicating the analysis process.

|  | Highest Values Recorded | Mean values | Lowest Values Recorded |
|---|---|---|---|
| Mean Elapsed Time (seconds) | 465.6 | 52.4 | 1.9 |
| Mean Percent of CPU used | 99.0 | 97.1 | 90.0 |
| Mean File system inputs | 1638291953.3 | 109563317.9 | 0.0 |
| Mean Involuntary context switches | 7875.5 | 1315.2 | 73.0 |
| Mean Voluntary Context Switches | 10455.5 | 1550.2 | 6.0 |
| Major I/O page faults | 540.0 | 101.7 | 0.0 |

Table I
DISTRIBUTION OF VALUES FOUND IN THE VOLATILITY EXPERIMENT

To obtain information regarding the file system structures within image files, the initial use of the `mmls` command was implemented. This command aimed to provide insights into the organization of data within these image files, facilitating a structured approach to further analysis [22]. However, during the examination, it was observed that `mmls` functioned effectively only with the BalenaRaspberry image file, offering a full description of its partition table.

Conversely, attempts to utilize `mmls` with other image files resulted in an error message stating *Cannot determine partition type*. This divergence in functionality restricted its effectiveness to solely the BalenaRaspberry image file, limiting the structured analysis to this specific OS.

The second command employed in the investigation was `blkls` from Sleuth Kit, chosen to access the image file and extract unallocated data blocks. This command utilizes various arguments, notably the sector offset number and file system type, which can be obtainable from the output generated by the `mmls` command. From Figure 6, multiple sectors can be observed, and the offset 8192 is used with the file system type FAT32.

Upon examination, it was discovered that the selected sector (8192) did not contain any unallocated data blocks, resulting in an empty output. This absence of data in the specified sector underscores the lack of potentially malicious content within the unallocated blocks at that particular location. However, this experimental process serves as evidence that had there been any malicious data present, 'blkls' would have been instrumental in detecting and extracting such data.

Key performance metrics for the operating systems that are similar to those listed in section IV-C are shown in the tables below. All of the operating systems have an average mean elapsed time of 0.0. `mainos.001` exhibits a notably higher average CPU usage of 82.0%, whereas Contiki shows a comparatively lower average utilization of 59.8%. All the others fell between 65.0% to 75.0%. Mean file system inputs represent the average volume of interactions with the file system during operations. FortiOS exhibits the highest average of 4120.0 file system inputs, while Balena displays the lowest at 78.0.

Figure 6. Balena Raspberry mmls and blkls commands

### E. Metadata Layer Analysis

In a traditional forensics scenario, there would be a continuation from the file system analysis onto metadata layer analysis, which is used to reconstruct individual files and attain file details or reconstruct deleted files.

Running the `fls` command provided an overview of the file hierarchy on each system. The results of this command are shown in Table X of the Appendix. Contiki showed the longest elapsed time with 0.5 seconds, yet yielded no positive results. FortiOS had the highest CPU usage of 85%, but the fastest time of only 0.01 seconds. Likewise it also yielded no positive results. The Windows IOT partitions performed very efficiently taking only 0.03 and 0.05 seconds for each of the two disks[6]. Windows IOT was able to return a positive file system output (seen in Figure 9). Balena performed in the middle of the group with 0.2 seconds of elapsed time and only 40% mean CPU usage, while also returning a readable file system that allowed for further analysis.

After the `fls` command was ran, then the `ils` command is performed on each operating system image. The results of this analysis are presented in Table XI. The commands run very quickly with elapsed time of 0.02 seconds, 0.02 seconds, 0.05 seconds, and 0.01 seconds, across the images for Windows IOT (mainos.001), FortiOS, Contiki, and Balena respectively. Similarly to the `fls` command, there were positive results from Windows IOT and Balena, however FortiOS and Contiki operations were unsuccessful.

The final test in the metadata layer analysis is to run the `istat` command on one of the inodes found in the earlier steps. In general, this command is a very a fast operation with the mean elapsed time equalling 0.01 seconds across both of the Windows IOT disks and 0.03 seconds when running on Balena. Both of these operating systems were successful in returning inode metadata. Inline with earlier results, the researchers were unable to get inode metadata using the `istat` command on either FortiOS or Contiki, both of which returned unsuccessful attempts after a mean elapsed time of 0.05 seconds each. The full results of `istat` analysis are presented in Table XII.

Across all of the tests, it was quickly apparent that two of the OS' were unrecognizable by the forensics tools selected for the study: FortiOS and Contiki. Any attempt to work with these images resulted in *unrecognizable file format* message from any Sleuth Kit commands that were executed. The researchers tried to run the tools by explicitly defining every supported file format from Sleuth Kit one-by-one, but returned zero positive results across all testing. This indicates a **full incompatibility** between these operating systems and Sleuth Kit when performing metadata layer analysis. Therefore, the results provided by this study when analyzing these two OS's didn't yield any worthwhile results.

The study was successful in demonstrating functional results with both Windows IOT and Balena OS systems. The research was able to successfully run `fls`, `ils`, and `istat` against both Windows IOT and Balena.

The file system of Balena OS resembled the familiar operating system hierarchy of a Linux operating system. The file format was `ext4`, which is something that Sleuth Kit has known direct compatibility with. Using an inode offset of 90112, it was possible to find and extract a cache file. Touching a file and then deleting it was able to find an allocated inode of the deleted file, making it possible to possibly recover a deleted file from a Balena operating device through the use of the `ils` command.

Similarly, the tests against Windows IOT worked well as Sleuth Kit recognized these images as similar to the core Windows Operating System. The file format was `ntfs`, just like the core Windows OS and the Sleuth Kit commands were able to navigate files and find inodes.

Overall performance of metadata layer commands was remarkably fast, with minimal impact to the investigation device. The mean duration for the commands across all operating systems was 0.068 seconds for the `fls` commands, 0.04 seconds for the `ils` commands, and 0.03 seconds for the `istat` commands. This means that all of these commands run extremely quickly and performance is not a major concern or factor when considering these types of forensics tasks.

## V. DISCUSSION

The discussion section delves into the implications and insights drawn from the empirical findings, offering insights into broader implications and potential avenues for future research.

### A. IoT devices and digital forensics

The proliferation of IoT devices has become widespread, yet the field of digital forensics still lacks widely adopted and robust tools capable of effectively managing these diverse devices. It is clear that traditional forensics tools are not up to the task to effectively be utilized on IOT Operating systems directly. There may be more success with interfacing with these IOT systems once they are fully integrated into a product, however, when interfacing directly with the operating system, the issues become very clear.

---

[6]Windows was split into two disks, Mainos.001 and Data-F

### B. Successes of Balena Raspberry

During the evaluation of the Balena Raspberry, several observations and efforts were undertaken to understand the complexities of its file system and the forensic analysis tools linked with it. The study brought up concerns about the file system format, particularly the comparison between FAT32 and ext2/ext3 formats, which could potentially affect the results of the forensic examination. A significant observation that emerged during the analysis was that only the Balena Raspberry could run `mmls` successfully. All the other files gave an error message saying "Cannot determine partition type." This discrepancy raises the hypothesis that the disk image utilized might be logical rather than physical.

Running `mmls` on the Balena Raspberry image file successfully gave a offset number, which could not be obtained with the other images being studied. This offset number served as a vital point of reference for subsequent analysis. During the experiment, different Sleuth Kit tools like `ils`, `fsstat`, and `fls` were used to look into the BalenaRaspberry image's file system. `fsstat` primarily displayed indications of End-of-File (EOF) markers as seen in Figure 8, The `fls` command unveiled a subset of filenames and associated information shown in Figure 7, whereas `ils` with specified offsets and file system types attempted to retrieve metadata.

### C. Successes of Windows IoT

The second operating system that the study was successful in performing forensics tasks against was the Windows IoT Operating System. This operating system follows a format almost identical to Windows, albeit extremely stripped down from a full desktop Windows installation.

The study found that common open-source tools that work well on desktop Windows systems, like FTK Imager and Sleuth Kit, also performed well against the Windows IoT variant. The only exception to this observation was the lack of proper volatile memory rendered Volatility as a tool largely useless for investigations, as discussed in Section III-B3.

FTK Imager performed efficiently when generating the Windows IoT images. Likewise, successful commands were recorded when performing file system and metadata layer analysis using Sleuth Kit commands `blkls`, `fls`, `ils`, and `istat`. The hypothesis being that the similarities in file format and file structure allowed Sleuth Kit to work with Windows IoT based on its existing compatibility with Windows, without the need for special tuning or customizations.

Study researchers noted a familiarity in the directory hierarchy within this image. This includes files on the main operating system partition such as `ProgramFiles`, `SystemData`, and `$OrphanFiles`; as demonstrated in Figure 9. The familiar `ntfs` file system allows investigators to use many known exploits for recovering hidden files on this IoT operating system [17]. Digital forensics investigators whom are

Figure 7. Balena Raspberry fls command

already familiar with collecting evidence on Windows operating systems will have no trouble collecting evidence from Windows IOT devices using their existing tools, techniques, and strategies.

## VI. LIMITATIONS

This study did maintain a few limitations as part of the methodology. The most notable limitation identified by the researchers is that the study was performed on clean devices, which did not contain *wear and tear*. For example, a fresh standard install of each operating system did not contain any deleted files for the research to investigate.

Another known limitation is that performing the same actions on a specific IoT device may differ slightly from the experience of researchers on this study since a specific device implementation of an operating system may alter the file system slightly from what was experienced in the study. Each operating system represents a starting point for the IoT device hosting that operating system, but customizations, sometimes significant, may be performed on individual devices in accordance to each device manufacturer. Therefore the exact experience of the researchers may not represent the exact experience in a real world forensic scenario.

```
sansforensics@siftworkstation: ~/Desktop
$ fsstat -o 8192 BalenaRaspberry.001
FILE SYSTEM INFORMATION
--------------------------------------------
File System Type: FAT32

OEM Name: mkfs.fat
Volume ID: 0x3ac2a952
Volume Label (Boot Sector): resin-boot
Volume Label (Root Directory): resin-boot
File System Type Label: FAT32
Next Free Sector (FS Info): 16304
Free Sector Count (FS Info): 65615

Sectors before file system: 0

File System Layout (in sectors)
Total Range: 0 - 81919
* Reserved: 0 - 31
** Boot Sector: 0
** FS Info Sector: 1
** Backup Boot Sector: 6
* FAT 0: 32 - 661
* FAT 1: 662 - 1291
* Data Area: 1292 - 81919
** Cluster Area: 1292 - 81919
*** Root Directory: 1292 - 16304

METADATA INFORMATION
--------------------------------------------
Range: 2 - 1290054
Root Directory: 2

CONTENT INFORMATION
--------------------------------------------
Sector Size: 512
Cluster Size: 512
Total Cluster Range: 2 - 80629

FAT CONTENTS (in sectors)
--------------------------------------------
1292-1292 (1) -> 16302
1293-1293 (1) -> EOF
1294-1331 (38) -> EOF
1332-1434 (103) -> EOF
1435-1537 (103) -> EOF
1538-1641 (104) -> EOF
```

Figure 8. Balena Raspberry fsstat command

```
sansforensics@siftworkstation: ~/Downloads
$ fls mainos.001
r/r 4-128-4:    $AttrDef
r/r 8-128-2:    $BadClus
r/r 8-128-1:    $BadClus:$Bad
r/r 6-128-4:    $Bitmap
r/r 7-128-1:    $Boot
d/d 11-144-4:   $Extend
r/r 2-128-1:    $LogFile
r/r 0-128-6:    $MFT
r/r 1-128-1:    $MFTMirr
r/r 9-128-8:    $Secure:$SDS
r/r 9-144-11:   $Secure:$SDH
r/r 9-144-14:   $Secure:$SII
r/r 10-128-1:   $UpCase
r/r 10-128-4:   $UpCase:$Info
r/r 3-128-3:    $Volume
d/d 36-144-1:   Data
d/d 8471-144-1: Program Files
d/d 8478-144-1: ProgramData
d/d 45-144-1:   PROGRAMS
d/d 37-144-1:   System Volume Information
d/d 8541-144-1: SystemData
d/d 42-144-1:   Users
d/d 39-144-5:   Windows
V/V 12800:      $OrphanFiles
sansforensics@siftworkstation: ~/Downloads
$
```

Figure 9. File structure of Windows IOT resembles the primary Windows Operating System

The study's methodology leveraged the use of virtual machines to extract data from simulated operating systems. It is noted that this is not representative of many real-world scenarios that would require data extraction from a live device.

The IoT space is expansive and many additional operating systems still exist beyond the small subset selected for the study. Every attempt was made to select popular IoT operating systems but many others exist which were not part of the study. Lastly, it is noted that the study focused on using open source, easily available forensics tools for the analysis. This is not exhaustive of all tools available to forensics investigators, especially many premium products are available.

## VII. Future Research

The goal of this research was to fill known gaps in the limited research space surrounding forensic investigation of IoT devices. Although, the researchers of this study acknowledge that despite this contribution, much additional research should still be continued to further explore this area.

This study focused specifically on using the Open Source forensic investigation tools on IoT device operating systems, but it would be ideal to expand the research to a wide array of additional tools, particularly premium products offered from private organizations. Likewise, this same study methodology could be expanded through future research to include additional IoT operating systems, since many others have been identified. Similarly the same methodology could be used towards targeting the study of specific devices, instead of the operating systems.

The research and production of new methodologies for extracting information directly from IoT devices would be an additional area of significant benefit to the industry. Current data extraction methods are limited due to a lack of valid inputs on devices or a lack of direct OS-level access to devices. Creative methodologies for data extraction of IoT devices should be further explored.

The most important takeaway that the researchers would like to encourage is the creation of new forensics tools that are designed specifically to handle IoT devices. This could also be performed by the creation of new plugins, extensions, or improvements to the current suite of tools (like a Volatility plugin or new file system types added in Sleuth Kit). The lack of compatibility between the current suite of forensic tools and popular IoT OS's hopes to encourage the demand and industry need for tailor designed tools to address this digital forensics weak spot.

## VIII. CONCLUSION

This research evaluated the efficiency and speed of open-source forensic tools in collecting digital evidence from various IoT devices. By rigorously testing these tools across multiple IoT operating systems, the study provides valuable insights into their performance in common digital forensics tasks.

The meticulous recording and analysis of time spent on each task have allowed for a comprehensive assessment of each tool's effectiveness. Furthermore, the study offers practical recommendations for IoT security experts and digital forensic practitioners, focusing on the most effective open-source tools for forensic investigations.

The study tested these forensics tasks against four IoT Operating Systems: Balena OS, Windows IOT, Contiki, and FortiOS. Of these, only Balena OS and Windows 'IoT, were found to be practical for evidence collection with standard open-source forensics tools.

This investigation underscores a discernible deficiency in the availability of proficient open-source tools for the forensic examination of IoT devices. The findings of this research underscore the necessity for the development of specialized tools tailored for these devices and an enhanced awareness of evidence collection methodologies pertinent to IoT environments.

## APPENDIX

| Volatility imageInfo | mainos.001 | FortiOS | EFI | Data-f | Contiki | Balena |
|---|---|---|---|---|---|---|
| Mean Elapsed Time (seconds) | 420.3 | 23.0 | 22.0 | 153.2 | 465.6 | 382.5 |
| Mean Percent of CPU used | 99.0 | 98.5 | 99.0 | 99.0 | 99.0 | 99.0 |
| Mean File system inputs | 562851488.0 | 137468.0 | 131320.0 | 250172792.0 | 425392693.3 | 367395544.0 |
| Mean Involuntary context switches | 5446.0 | 1449.0 | 4840.0 | 1392.0 | 7875.5 | 6917.0 |
| Mean Voluntary Context Switches | 4875.0 | 576.5 | 17.0 | 1408.0 | 10455.5 | 5114.0 |
| Major I/O page faults | 534.0 | 41.5 | 4.0 | 152.0 | 470.8 | 540.0 |

Table II
VOLATILITY IMAGEINFO COMMAND RESULTS ON IOT OPERATING SYSTEMS

| Volatility pstree | mainos.001 | FortiOS | EFI | Data-f | Contiki | Balena |
|---|---|---|---|---|---|---|
| Mean Elapsed Time (seconds) | 7.8 | 2.4 | 2.3 | 3.6 | 6.6 | 5.7 |
| Mean Percent of CPU used | 96.0 | 95.0 | 90.0 | 92.0 | 93.2 | 94.0 |
| Mean File system inputs | 8643584.0 | 38552.0 | 46520.0 | 575096.0 | 5269557.3 | 2662496.0 |
| Mean Involuntary context switches | 497.0 | 372.0 | 430.0 | 564.0 | 1315.3 | 540.0 |
| Mean Voluntary Context Switches | 2722.0 | 1001.0 | 1958.0 | 2341.0 | 2445.0 | 2613.0 |
| Major I/O page faults | 128.0 | 55.0 | 55.0 | 97.0 | 133.2 | 124.0 |

Table III
VOLATILITY PSTREE COMMAND RESULTS ON IOT OPERATING SYSTEMS

| Volatility connscan | mainos.001 | FortiOS | EFI | Data-f | Contiki | Balena |
|---|---|---|---|---|---|---|
| Mean Elapsed Time (seconds) | 7.5 | 2.3 | 1.9 | 3.1 | 6.8 | 4.5 |
| Mean Percent of CPU used | 98.0 | 99.0 | 98.0 | 98.0 | 97.0 | 94.0 |
| Mean File system inputs | 8529240.0 | 0.0 | 0.0 | 0.0 | 4904713.3 | 176816.0 |
| Mean Involuntary context switches | 113.0 | 145.0 | 500.0 | 773.0 | 689.2 | 289.0 |
| Mean Voluntary Context Switches | 885.0 | 10.0 | 7.0 | 7.0 | 1299.0 | 2410.0 |
| Major I/O page faults | 99.0 | 0.0 | 0.0 | 0.0 | 92.2 | 68.0 |

Table IV
VOLATILITY CONNSCAN COMMAND RESULTS ON IOT OPERATING SYSTEMS

| Volatility pslist | mainos.001 | FortiOS | EFI | Data-f | Contiki | Balena |
|---|---|---|---|---|---|---|
| Mean Elapsed Time (seconds) | 7.6 | 2.2 | 2.0 | 3.1 | 7.0 | 4.3 |
| Mean Percent of CPU used | 98.0 | 99.0 | 97.0 | 98.0 | 97.5 | 99.0 |
| Mean File system inputs | 8447832.0 | 0.0 | 0.0 | 0.0 | 3054936.0 | 73128.0 |
| Mean Involuntary context switches | 120.0 | 126.0 | 830.0 | 806.0 | 622.5 | 152.0 |
| Mean Voluntary Context Switches | 831.0 | 7.5 | 6.0 | 13.0 | 1248.8 | 99.0 |
| Major I/O page faults | 101.0 | 0.0 | 0.0 | 0.0 | 91.2 | 2.0 |

Table V
VOLATILITY PSLIST COMMAND RESULTS ON IOT OPERATING SYSTEMS

| Volatility hivescan | mainos.001 | FortiOS | EFI | Data-f | Contiki | Balena |
|---|---|---|---|---|---|---|
| Mean Elapsed Time (seconds) | 8.0 | 2.2 | 1.9 | 3.0 | 6.9 | 4.0 |
| Mean Percent of CPU used | 96.0 | 99.0 | 98.0 | 99.0 | 97.2 | 99.0 |
| Mean File system inputs | 51904.0 | 0.0 | 0.0 | 0.0 | 1638291953.3 | 51904.0 |
| Mean Involuntary context switches | 501.0 | 132.5 | 489.0 | 752.0 | 704.8 | 73.0 |
| Mean Voluntary Context Switches | 2362.0 | 8.0 | 10.0 | 12.0 | 1539.0 | 227.0 |
| Major I/O page faults | 106.0 | 0.0 | 0.0 | 0.0 | 91.7 | 66.0 |

Table VI
VOLATILITY HIVESCAN COMMAND RESULTS ON IOT OPERATING SYSTEMS

| | Fortinet Disk1 | Fortinet Disk2 | Balena | Win-IoT Disk1 | Win-IoT Disk2 | Win-IoT Disk3 |
|---|---|---|---|---|---|---|
| **Disks Segment** | FG0s2 | FG0s3 | BalenaRaspberry.001 | Data-f.002 | efi-e.001 | mainos |
| **Bytes per Sector** | 512 | 512 | 512 | 512 | 512 | 512 |
| **Image Type** | VMWare Virtual Disk | VMWare Virtual Disk | Raw (dd) | Virtual Disk | Virtual Disk | Virtual Disk |
| **Source Data Size** | 2048 MB | 30720 MB | 908 MB | 2095 MB | 64 MB | 1428 MB |
| **Sector Count** | 4194304 | 62914560 | 1859584 | 4292064 | 131072 | 2926584 |
| **Acquisition Time** | 4s | 1m 1s | 6s | 2m16s | 4s | 1m52s |

Table VII
COMPARISON OF IMAGE ACQUISITION PARAMETERS BY FTK IMAGER

Table VIII
PERFORMANCE METRICS OF 'MMLS' COMMAND

|  | mainos.001 | FortiOS | EFI | Data-f | Contiki | Balena |
|---|---|---|---|---|---|---|
| Mean Elapsed Time (seconds) | 0.0 | 0.1 | 0.0 | 0.0 | 0.0 | 0.0 |
| Mean Percent of CPU used | 82.0 | 65.5 | 75.0 | 65.0 | 59.8 | 66.0 |
| Mean File system inputs | 536.0 | 4120.0 | 528.0 | 3904.0 | 548.0 | 78.0 |
| Mean Involuntary context switches | 19.0 | 22.0 | 14.0 | 4.0 | 24.3 | 19.0 |
| Mean Voluntary Context Switches | 2.0 | 20.5 | 2.0 | 41.0 | 5.7 | 4.0 |
| Major I/O page faults | 0.0 | 19.5 | 0.0 | 41.0 | 0.0 | 0.0 |

Table IX
DISTRIBUTION OF VALUES IN FILE SYSTEM ANALYSIS EXPERIMENT

|  | Highest Values Recorded | Mean values | Lowest Values Recorded |
|---|---|---|---|
| Mean Elapsed Time (seconds) | 0.1 | 0.0 | 0.0 |
| Mean Percent of CPU used | 82.0 | 68.9 | 59.8 |
| Mean File system inputs | 4120.0 | 1619.0 | 78.0 |
| Mean Involuntary context switches | 24.3 | 17.1 | 4.0 |
| Mean Voluntary Context Switches | 41.0 | 12.5 | 2.0 |
| Major I/O page faults | 41.0 | 10.1 | 0.0 |

|  | mainos.001 | FortiOS | Data-f | Contiki | Balena |
|---|---|---|---|---|---|
| Mean Elapsed Time (seconds) | 0.03 | 0.01 | 0.05 | 0.50 | 0.20 |
| Mean Percent of CPU used | 73.0 | 85.0 | 42.0 | 50.0 | 40.0 |
| Mean File system inputs | 1688.0 | 0.0 | 18944.0 | 3984.0 | 272.0 |
| Mean Involuntary context switches | 36.0 | 8.0 | 13.0 | 27.0 | 6.0 |
| Mean Voluntary Context Switches | 10.0 | 1.0 | 47.0 | 136.0 | 3.0 |
| Major I/O page faults | 10.0 | 0.0 | 0.0 | 2.0 | 0.0 |

Table X
FLS COMMAND ON IOT OPERATING SYSTEMS

|  | mainos.001 | FortiOS | Data-f | Contiki | Balena |
|---|---|---|---|---|---|
| Mean Elapsed Time (seconds) | 0.02 | 0.02 | 0.10 | 0.05 | 0.01 |
| Mean Percent of CPU used | 85.0 | 52.0 | 55.0 | 32.0 | 90 |
| Mean File system inputs | 5208.0 | 144.0 | 0.0 | 672.0 | 0.0 |
| Mean Involuntary context switches | 39.0 | 5.0 | 31.0 | 21.0 | 10.0 |
| Mean Voluntary Context Switches | 9.0 | 3.0 | 1.0 | 5.0 | 1.0 |
| Major I/O page faults | 9.0 | 2.0 | 0.0 | 5.0 | 0.0 |

Table XI
ILS COMMAND ON IOT OPERATING SYSTEMS

|  | mainos.001 | FortiOS | Data-f | Contiki | Balena |
|---|---|---|---|---|---|
| Mean Elapsed Time (seconds) | 0.01 | 0.05 | 0.01 | 0.05 | 0.03 |
| Mean Percent of CPU used | 78.0 | 44.0 | 75.0 | 50.0 | 50.0 |
| Mean File system inputs | 2313.0 | 13544.0 | 128.0 | 18996.0 | 9000.0 |
| Mean Involuntary context switches | 9.0 | 18.0 | 7.0 | 25.0 | 8.0 |
| Mean Voluntary Context Switches | 15.0 | 51.0 | 2.0 | 44.0 | 39.0 |
| Major I/O page faults | 14.0 | 3.0 | 0.0 | 0.0 | 0.0 |

Table XII
ISTAT COMMAND ON IOT OPERATING SYSTEMS